\documentclass[twocolumn,preprintnumbers,a4paper,
superscriptaddress,floatfix,twoside,prd]{revtex4}

\usepackage{bm}
\usepackage{epsfig}
\usepackage{graphics}
\usepackage{natbib}
\usepackage{fancyh}
\usepackage[l]{floatflt}
\usepackage{epsfig}
\usepackage{amssymb}
\usepackage{latexsym}
\usepackage{times}
\usepackage{slashed}
\usepackage{upgreek}
\usepackage{amsmath}
\usepackage{amsfonts}
\usepackage{amsbsy}
\usepackage{amscd}
\usepackage{bbm}

\newcommand{\Eref}[1]{Eq.~(\ref{#1})}
\newcommand{\Erefs}[2]{Eqs.~(\ref{#1}) -- (\ref{#2})}
\newcommand{\Sref}[1]{Sec.~\ref{#1}}
\newcommand{\Fref}[1]{Fig.~\ref{#1}}
\newcommand{\Tref}[1]{Table~\ref{#1}}
\newcommand{\cref}[1]{Ref.~\cite{#1}}

\newcommand{\hepth}[1]{{\ftn \tt hep-th/#1}}
\newcommand{\hepph}[1]{{\ftn\tt hep-ph/#1}}

\newcommand{\arxiv}[1]{{\ftn\tt  arXiv:#1}}

\newcommand{\bal}{\begin{align}}
\newcommand{\eall}{\end{align}}
\newcommand{\beqs}{\begin{subequations}}
\newcommand{\eeqs}{\end{subequations}}
\newcommand{\eec}{\end{center}}
\newcommand{\bec}{\begin{center}}

\newcommand{\eem}{\end{matrix}}
\newcommand{\bem}{\begin{matrix}}
\newcommand{\eeq}{\end{equation}}
\newcommand{\beq}{\begin{equation}}
\newcommand{\ba}{\begin{array}}
\newcommand{\ea}{\end{array}}
\newcommand{\bea}{\begin{eqnarray}}
\newcommand{\eea}{\end{eqnarray}}
\newcommand{\baq}{\begin{eqnarray}}
\newcommand{\eaq}{\end{eqnarray}}
\newcommand{\bel}{\begin{align}}
\def\eel{\end{align}}

\newcommand\eqs[2]{Eqs.~(\ref{#1}) and (\ref{#2})}

\newcommand\eqss[3]{Eqs.~(\ref{#1}), (\ref{#2}) and (\ref{#3})}

\newcommand{\ftn}{\footnotesize}

\newcommand{\TeV}{{\mbox{\rm TeV}}}

\newcommand{\GeV}{{\mbox{\rm GeV}}}

\newcommand{\sEref}[2]{Eq.~(\ref{#1}{\ftn\sf {#2}})}

\newcommand{\etal}{{\it et al.\/}}

\def\to{\rightarrow}

\def\lf{\left(}
\def\rg{\right)}
\newcommand\vev[1]{\left\langle {#1} \right\rangle}
\newcommand\veva[1]{\langle {#1} \rangle}
\newcommand\vevg{\langle {\rm g}\rangle}
\newcommand{\Gr}{\ensuremath{\widetilde{G}}}

\newcommand{\Vhi}{\ensuremath{V_{\rm H}}}
\newcommand{\Vhc}{\ensuremath{V_{\rm Hc}}}

\newcommand{\khi}{\ensuremath{K_{\rm H}}}
\newcommand{\khc}{\ensuremath{K_{\rm Hc}}}
\newcommand{\khio}{\ensuremath{K_{\rm H0}}}
\newcommand{\whi}{\ensuremath{W_{\rm H}}}
\newcommand{\wkhi}{\ensuremath{\wtilde K}}

\newcommand{\uc}{\ensuremath{u_{\rm c}}}
\newcommand{\vc}{\ensuremath{v_{\rm c}}}
\newcommand{\wc}{\ensuremath{w_{\rm c}}}

\newcommand{\Vhio}{\ensuremath{V_{\rm H0}}}
\newcommand{\Vhoc}{\ensuremath{V_{\rm H0c}}}

\newcommand{\mP}{\ensuremath{m_{\rm P}}}

\newcommand{\lm}{\ensuremath{\lambda_\mu}}

\def\openone{\leavevmode\hbox{\small1\kern-3.8pt\normalsize1}}

\newcommand{\dK}{\ensuremath{\Delta K_\mu}}
\newcommand{\km}{\ensuremath{C_H}}

\newcommand{\hd}{{\ensuremath{H_d}}}
\newcommand{\hu}{{\ensuremath{H_u}}}

\newcommand{\zm}{\ensuremath{Z_{-}}}
\newcommand{\zp}{\ensuremath{Z_{+}}}

\newcommand{\nm}{\ensuremath{\nu_{-}}}
\newcommand{\np}{\ensuremath{\nu_{+}}}
\newcommand{\no}{\ensuremath{N_{\rm c}}}

\newcommand{\what}{\ensuremath{\widehat}}
\newcommand{\wtilde}{\ensuremath{\widetilde}}

\def\bz{{\bar Z}}
\def\al{{\alpha}}

\def\bt{{\beta}}

\def\n{\bar{n}}

\def\th{{\theta}}

\newcommand{\mgr}{\ensuremath{m_{3/2}}}
\newcommand{\mgro}{\ensuremath{m_{3/2}}}
\newcommand{\mz}{\ensuremath{m_{\widehat{z}}}}
\newcommand{\mth}{\ensuremath{m_{\widehat\theta}}}
\newcommand{\mzo}{\ensuremath{m_{\widehat{z}}}}

\newcommand{\phc}{\ensuremath{\Phi}}

\def\Ka{K\"{a}hler potential}
\def\Km{K\"{a}hler manifold}

\newcommand{\im}{\ensuremath{{\sf Im}}}

\renewcommand{\refname}{{\bf\scshape References}}

\renewcommand{\thesubsection}{{\small\sf\Alph{subsection}}}

\renewenvironment{subequations}{%
\refstepcounter{equation}%
\setcounter{parentequation}{\value{equation}}%
  \setcounter{equation}{0}
  \def\theequation{\theparentequation{\sf\small\alph{equation}}}%
  \ignorespaces
}{%
  \setcounter{equation}{\value{parentequation}}%
  \ignorespacesafterend
}

\begin{document}

%%%%% ME DIOR8WSEIS META THN YPOBOLH STO PRD

\title{\bf\scshape  SUSY-Breaking Scenarios With a Mildly Violated {\boldmath $R$} Symmetry}

\author{{\scshape Constantinos Pallis} \\ {\small\it Laboratory of Physics, Faculty of
Engineering, Aristotle University of Thessaloniki, GR-541 24
Thessaloniki, GREECE} \\ {\ftn\sl  e-mail address: }{\ftn\tt
kpallis@gen.auth.gr}}

\begin{abstract}

\noindent {\ftn \bf\scshape Abstract:} New realizations of the
gravity-mediated SUSY breaking are presented consistently with an
$R$ symmetry. We employ monomial superpotential terms for the
hidden-sector (goldstino) superfield and \Ka s parameterizing
compact or non-compact \Km s. Their scalar curvature may be
systematically related to the $R$ charge of the goldstino so that
Minkowski solutions without fine tuning are achieved. A mild
violation of the $R$ symmetry by a higher order term in the \Ka s
allows for phenomenologically acceptable masses for the $R$ axion.
In all cases, non-vanishing soft SUSY-breaking parameters are
obtained and a solution to the $\mu$ problem of MSSM may be
accommodated by conveniently applying the Giudice-Masiero
mechanism.
\\ \\ {\scriptsize {\sf PACs numbers: 12.60.Jv, 04.65.+e}
%11.30.Er, 11.30.Pb,

%\hfill {\sl\bfseries Published in} {\sl Phys. Rev. D} {\bf 86},
%023523 (2012)

}
%\pacs{98.80.Cq, 11.30.Qc, 12.60.Jv} it does not require fine tuned parameters

\end{abstract}\pagestyle{fancyplain}

\maketitle

\rhead[\fancyplain{}{ \bf \thepage}]{\fancyplain{}{\sl
SUSY-Breaking Scenarios With a Mildly Violated $R$ Symmetry}}
\lhead[\fancyplain{}{\sl C. Pallis}]{\fancyplain{}{\bf \thepage}}
\cfoot{}

\section{Introduction}\label{intro}

The fact that \emph{Supersymmetry} ({\ftn\sf SUSY}) remains still
unobservable in the nature does not invalidate its importance in
constructing theories beyond the \emph{Standard Model} ({\ftn\sf
SM}). Most notably, the problem of SUSY breaking is currently
under intense scrutiny -- see e.g.
\cref{susyr,noscale18,noscaleinfl,ant,riotto,kai,buch,ketov,linde1,king,
nevz,Rgauged,burgess} -- and new settings are suggested within
\emph{Supergravity} ({\ftn\sf SUGRA}) \cite{nilles} aiming to
obtain natural Minkowski and/or de Sitter vacua -- the latter
possibility is not so favored, though, because of the controversy
\cite{vafa, lindev} surrounding this kind of (meta) stable vacua.
A crucial question in this area of research concerns
\cite{noscale18} the specific forms of superpotential and the type
of geometry which may be adopted.

Trying to reply to these questions in a recent paper \cite{susyr},
we employed as guideline an approximate $R$ symmetry \cite{nilles,
hall} which totally fixes the form of the superpotential, $\whi$,
as a function of the goldstino superfield, $Z$, -- contrary to the
well-known and widely adopted Polonyi model \cite{polonyi} where
the same $R$ symmetry is badly violated by a constant term in
$\whi$. We show that if we associate $\whi$ with a \Ka, $\khi$,
which parameterizes the $SU(1,1)/U(1)$ \Km\ with constant
curvature $-1/2$ we can obtain a Minkowski vacuum (i.e., SUSY
breaking with vanishing classical vacuum energy) without unnatural
fine tuning, and non-vanishing \emph{soft SUSY-breaking} ({\sf\ftn
SSB}) parameters \cite{soft,kapnilles}, of the order of the
gravitino mass, $\mgr$, which is readily determined at the tree
level. A solution to the $\mu$ problem of \emph{Minimal
Supersymmetric SM} ({\ftn\sf MSSM}) may be also achieved by
suitably applying the Giudice-Masiero mechanism \cite{masiero} --
for an updated review see \cref{mubaer,baer}. The possibly
problematic, from astrophysical point of view \cite{astro},
(pseudo) Nambu-Goldstone boson named $R$ axion \cite{rnelson} may
acquire acceptably large mass  by mildly violating the $R$
symmetry in $\khi$ without any impact either to the minimization
of the SUGRA potential or to the values of the SSB parameters.

In the present paper, motivated by several related models
\cite{qbinetruy, fractional, nevz, noscale18, castano}, we extend
our approach above, considering different $R$ charges for $Z$ and
allowing, thereby, for other (not only linear) monomial forms of
$\whi(Z)$ -- see \Sref{hd}. We show that natural Minkowski
solutions, similar to those found in \cref{susyr}, can be
accommodated by conveniently selecting the geometry of the
internal space. Contrary to \cref{susyr}, we here consider compact
\cite{su11, pana} and non-compact \cite{linde, noscale18} \Km s,
whose the curvature is determined as a simple function of the $R$
charge of $Z$. In both cases, we specify new SUSY-breaking minima
and show that the same quartic term introduced in \cref{noscale18,
susyr} provides mass for the $R$ axion without spoiling, though,
the minimization of the SUGRA potential. Following the analysis of
\cref{susyr}, we combine our hidden sectors with some sample
observable ones and broaden our results as regards the SSB
parameters and the derivation of the $\mu$ term of MSSM -- see
\Sref{obs}. Our conclusions are further discussed in \Sref{con}.
An alternative formulation of our set-up is presented in Appendix
A.

Unless otherwise stated, we use units where the reduced Planck
mass $\mP=2.433\cdot 10^{18}~\GeV$ is taken to be unity and charge
conjugation is denoted by a bar.

% -- cf. \cref{dvali, univ}To some extent,
%Although quite compelling, the above scheme gets into trouble due
%to the presence of the massless $R$ axion.

\section{Hidden Sector} \label{hd}

Here we first -- in \Sref{hd1} -- specify the hidden sector of our
model and then -- in \Sref{hd2} --  investigate the SUSY-breaking
mechanism employing the curvature of the \Km\ as free parameter.
Lastly, in \Sref{hd3}, we introduce the $R$-symmetry violating
term and compute the mass of the $R$ axion.

\subsection{\sc\small\sffamily  Model Set-up}\label{hd1}

Following our strategy in \cref{susyr}, we consider a hidden
sector consisting of just one gauge-singlet superfield $Z$. The
corresponding superpotential $W_{\rm H}$ is fixed by imposing an
$R$ symmetry under which $Z$ has $R$ charge $2/\nu$ -- and not
just $2$ as in \cref{susyr} -- whereas that of $\whi$ is taken to
be $2$, i.e.,
\beq W_{\rm H} = m Z^\nu, \label{whi} \eeq
where $m$ is a positive, free parameter with mass dimensions. Also
$\nu\neq0$ is an exponent which may, in principle, acquire any
real value. This is possible if we consider $\whi$ as an effective
superpotential valid close to the non-zero vacuum of the theory.
E.g., we can obtain $\nu<0$ from dynamical breaking of a $SU(N)$
continuous symmetry \cite{dine} assuming in addition
\cite{qbinetruy} that this is not a gauge symmetry to avoid D-term
contributions in the SUGRA potential. Moreover, $\nu>0$ fractional
numbers are already considered in \cref{castano, noscale18} and,
as discussed in Appendix A, support a complete form of $\whi$
invoking an appropriate field redefinition \cite{fractional}.

Contrary to the Polonyi model \cite{polonyi} we do not consider
any $R$-symmetry violating constant term in $\whi$. Such a
violation is accommodated in the \Ka\ which is
\beq K_{\rm H}=N\ln\lf1+\frac{|Z|^2-k\zm^4}{N}\rg\label{khi} \eeq
with $Z_\pm=Z\pm\bz$. Here $N$ and $k$ are free parameters of
either sign. Motivated by several superstring and D-brane models
\cite{ibanez} we consider the integer negative values of $N$ as
the most natural. However, positive $N$'s can not be disregarded
if we limit our attention on SUGRA \cite{su11, pana}. On the other
hand, $k$ is a parameter which mildly violates $R$ symmetry. We
call this violation ``mild'' since {\sl (a)} $k$ is involved in a
higher-order (non quadratic) term in $K_{\rm H}$ and {\sl (b)}
tiny $k$ values endow $R$ axion with phenomenologically acceptable
masses -- see \Sref{hd3}. The presence of $k$ in $K_{\rm H}$ is
totally natural, according to the argument \cite{symm} of 't
Hooft, since nullifying it the $R$ symmetry becomes exact. In view
of this fact, the positivity of the argument of logarithm in
\Eref{khi} implies \beq \label{logbound1} 1+|Z|^2/N
\gtrsim0~~\Rightarrow~~-|Z|^2/N \lesssim1\eeq which is true for
any $N>0$ but dictates  \beq |Z|^2\lesssim
-N~~\mbox{for}~~N<0\,.\label{logbound}\eeq

The curved space parameterized by $\khi$ has metric
\beqs\beq{\rm g}:=\partial_Z\partial_\bz \khi=Nw\lf
N+|Z|^2-k\zm^4\rg^{-2}\label{ds}\eeq
where we find it convenient to define \beq \label{w}
w=N(1+12k\zm^2)+k\zm^2(3\zp^2+4k\zm^4).\eeq\eeqs
%N+4k^2\zm^{6}+3k\zm^{2}\big(\zm^2+4(|Z|^2+N)\big)

Substituting \eqs{whi}{khi} in the well-known formula
\beq \label{Vsugra} V:= e^{G}\lf G^{I\bar J} G_I G_{\bar
J}-3\rg~~\mbox{with}~~G:= \khi + \ln |W_{\rm H}|^2,\eeq
$I,J=Z$, $G^{Z\bar Z}={\rm g}^{-1}$, $G_I:= \partial_I G$ and
$G_{\bar J}:=
\partial_{\bar J} G$, we find the hidden-sector (F-term) scalar potential
\beq\Vhi=\frac{m^2}{N}e^{\khi}|Z|^{2(\nu-1)}\lf
\frac{uv}{w}-3N|Z|^2\rg.\label{Vhzz}\eeq Here $w$ originates from
the numerator in \Eref{ds} and we introduce the quantities
\beqs\bel \label{u} u&=N\nu+(N+\nu)|Z|^2-k\zm^{3}\lf\nu\zm+4NZ\rg,\\
\label{v} v&=N\nu+(N+\nu)|Z|^2-k\zm^{3}\lf\nu\zm-4N\bz\rg,
\end{align}\eeqs\\ [-0.4cm]
which arise from the numerators of $G_Z$ and $G_{\bar Z}$ in
\Eref{Vsugra}. Written in this form, $\Vhi$ may be directly
compared to the one employed in \cref{susyr}.

\subsection{\sc\small\sffamily  The $R$-Symmetric Limit}\label{hd2}

If we set $k=0$ in \Eref{khi} we obtain the exactly $R$-symmetric
form of $\khi$, $\khio$, which parameterizes the coset spaces
$SU(1,1)/U(1)$ for $N<0$ or $SU(2)/U(1)$ for $N>0$ \cite{su11},
with metric and constant scalar curvature respectively
\beq\label{ds0} {\rm g}^{(0)}=\partial_Z\partial_\bz K_{\rm H0}
=\lf1+\frac{|Z|^2}{N}\rg^{-2}\mbox{and}~~{\mathcal R}_{\rm
H}^{(0)}=\frac2N\,.\eeq  The last quantity reveals that the \Km\
is compact or non-compact if $N>0$ or $N<0$. Here and hereafter,
the superscript $(0)$ and the subscript $0$ denote quantities
corresponding to the totally $R$-symmetric case. Working in this
simplified framework, it is more convenient to investigate the
existence of Minkowski solutions, as done in \Sref{hd21}, classify
the relevant solutions -- see \Sref{hd22} -- and derive the
relevant particle spectrum in \Sref{hd23}.

Let us clarify here that the $SU(1,1)/U(1)$ manifold is also
parameterized by the half-plane coordinates (usually notated by
$T$ and $\bar T$) which are related to the disc coordinates ($Z$
and $\bar Z$) through a Cayley transform \cite{linde, susyr,
su11}. This parameterization, though, violates the $R$ symmetry
and so it is inappropriate for our purposes. Note that the
parameterization of the $SU(2)/U(1)$ space does not change under
that transform.

\subsubsection{\small\sf Minkowski Vacua} \label{hd21}

Substituting $k=0$ into \Eref{Vhzz} and taking into account that
\beq \label{Gz}G^{(0)}_Z=\bar G^{(0)}_\bz =\frac1Z\lf\sqrt{{\rm
g}^{(0)}}|Z|^2+\nu\rg\,. \eeq
we derive the corresponding SUGRA potential
\beq \label{Vh}\Vhio
=\frac{m^2}{N^2}\frac{e^{\khio}}{|Z|^{2(1-\nu)}} \lf\lf\nu N + \lf
N+\nu\rg |Z|^2\rg^2-3N^2 |Z|^2\rg,\eeq
which, as expected, depends exclusively on $|Z|^2$. The systematic
search of a Minkowski vacuum -- defined by the simultaneous
fulfilment of the conditions
\beq \label{cond} \mbox{\sf\ftn
(a)}~~\vev{\Vhio}=0,~~\mbox{\sf\ftn
(b)}~~\vev{\Vhio'}=0~~~\mbox{and}~~~\mbox{\sf\ftn
(c)}~~\vev{\Vhio''}>0,\eeq
where the derivatives \emph{with respect to} ({\sf\ftn w.r.t})
$|Z|^2$ are denoted by a prime -- is facilitated if the expression
in the parenthesis of \Eref{Vh} is expanded as follows
\beq \lf2\nu(N+\nu)-3N\rg N|Z|^2+\nu^2N^2+|Z|^4(N+\nu)^2.
\label{exp}\eeq
From our experience in \cref{susyr}, we suspect that the
attainment of a Minkowski vacuum is, in principle, possible if the
expression in \Eref{exp} equals to a perfect square trinomial
including a minus in the double product. Since the last two terms
consist a sum of squares, the desired perfect square trinomial is
achieved, if the two first terms in \Eref{exp} coincide with the
minus double product of the square root of anyone of the two last
terms. More explicitly, we impose the condition
\beq -2|N||\nu||N+\nu|=\lf2\nu(N+\nu)-3N\rg N\label{condn}\eeq
which, fortunately, is independent of $|Z|^2$.  We seek below the
$N=\no$ values as a function of $\nu$ that satisfy the condition
above. It is obvious that we need a combination of $\nu$, $\no$
and $\no+\nu$ such that $-2|\no||\nu||\no+\nu|=-2\no\nu(\no+\nu)$
since, otherwise, this term is cancelled out in \Eref{condn} and
the only possible solution is $\no=0$. As a consequence, we single
out the following cases which yield $\no\neq0$:

\begin{itemize}

\item[{\ftn\sf (i)}] $\nu>0$ and $\no>0$ resulting to $\nu+\no>0$;
\item[{\ftn\sf (ii)}] $\nu<0$, $\no>0$ and $\nu+\no<0$;
\item[{\ftn\sf (iii)}] $\nu>0$, $\no<0$ and $\nu+\no<0$.

\end{itemize}

If we solve \Eref{condn} w.r.t $\no$, in all cases above, we can
derive $\no$ using $\nu$ as free variable. Namely,
\beq
\no=\frac{4\nu^2}{3-4\nu}~~\Rightarrow~~\no\gtrless0~~\mbox{for}~~\nu\lessgtr3/4\,.\label{sol}\eeq
Turning the argument around, we can obtain two possible solutions
$\nu_{\pm}$ for every (preferably integer) $\no$ value if we solve
\Eref{sol} w.r.t $\nu$, i.e.,
\beq
\nu_{\pm}=\frac12\lf-\no\pm\sqrt{\no(\no+3)}\rg\label{nupm}\eeq
and correspond to real numbers if
\bea \no(\no+3)\geq0~~\Rightarrow~~
\no\geq0&\mbox{or}~~~\no\leq-3.\label{Nbounds}\eea
%
%
%\begin{table}[t!]
%\caption{\normalfont }
\begin{table}[t] \caption{\normalfont Solutions to \Eref{sol} for various
$\nu$'s.}
\begin{ruledtabular}
\begin{tabular}{c|cccccc}
$\nm$&  $5/4$&$6/5$ & $1$& $7/8$&$5/6$  &$4/5$ \\
$\no$&  $-25/8$&$-16/5$ & $-4$&$-49/8$& $-25/3$ &$-64/5$
\\\hline
$\nm$& $-1/2$&$-1$ & $-3/2$& $-2$ &$-5/2$&$-3$ \\
$\no$&  $1/5$&$4/7$ & $1$& $16/11$ &$25/13$&$12/5$  \\
\end{tabular}
\end{ruledtabular} \label{tab1}
\end{table}

For $N=\no$, $\Vhio$ in \Eref{Vh} includes by construction the
square of a binomial with a minus inside. Indeed, it reads
\beq \label{Vh2}\Vhoc
=\lf\frac{m}{4\nu}\rg^2\lf1+\frac{|Z|^2}{\no}\rg^{\no}|Z|^{2(\nu-1)}
\lf 3 |Z|^2 -4\nu^2\rg^2.\eeq
Here and hereafter the subscript ``c'' denotes quantities computed
for $N=\no$. Enforcing \sEref{cond}{b} we find a honest extremum
of $\Vhoc$ as follows \beq
\vev{\Vhoc'}=0~\Rightarrow~\vev{|Z|^2}=4|\nu|^2/3.\label{vev} \eeq
This result is, initially, restricted to $\nu\leq3/2$ for $\no<0$
by virtue of \eqs{logbound}{Nbounds}. It can be further
constrained from the observation that \sEref{cond}{a} is fulfilled
for \beq\no\neq-3~\Leftrightarrow~\nu\neq3/2. \label{neq32}\eeq
Indeed, taking $\no=-3$ and $\nu=3/2$, $\Vhoc$ in \Eref{Vh2} is
simplified as \beq \Vhoc=27m^2\sqrt{|Z|^2}/(12-4|Z|^2),\eeq which
exhibits just a SUSY vacuum. Finally, in order to check the
validity of \sEref{cond}{c}, we compute
\beq  \label{Vhp}
\vev{\Vhoc''}=2^{2\nu-5}3^{3-\nu}|\nu|^{2(\nu-1)}m^2\vevg^{-\no/2},\eeq
where we introduce the quantity \beq\vevg=\veva{{\rm
g}^{(0)}}=\frac14\lf1-\frac23\nu\rg^{-2}=\frac{9\no}{16\nu^2(\no+3)},
\label{defg}\eeq which controls the stability of the vacuum in
\Eref{vev} -- as we show below, even for $k\neq0$ in \Eref{khi}
the Minkowski vacuum lies along the direction $Z=\zm$ and so we do
not apply the symbolic distinction mentioned below \Eref{ds0}
regarding $\vevg$ and $\veva{{\rm g}^{(0)}}$. Namely, $\nu<3/2$
assures the validity of \sEref{cond}{c} for any $\no$. Taking into
account also, the constraint below \Eref{vev} for $\no<0$ and
\eqs{sol}{Nbounds} we end up with the following allowed domains
\beq \label{all}
\frac34<\nu<\frac32~~\mbox{for}~~\no<0~~\mbox{and}~~\nu<\frac34~~\mbox{for}~~\no>0.\eeq
Note that the $\nu$ range is rather limited for $\no<0$ and ampler
for $\no>0$.

%%%%%%%%%%%%%%%%%%%%%%%%%%%%%%%%%%%%%%%%%%%%%%%%%%%%%%%%%%%%%%%%%%%%
\begin{figure}[!t]%\vspace*{-.25in}
\includegraphics[width=60mm,angle=-90]{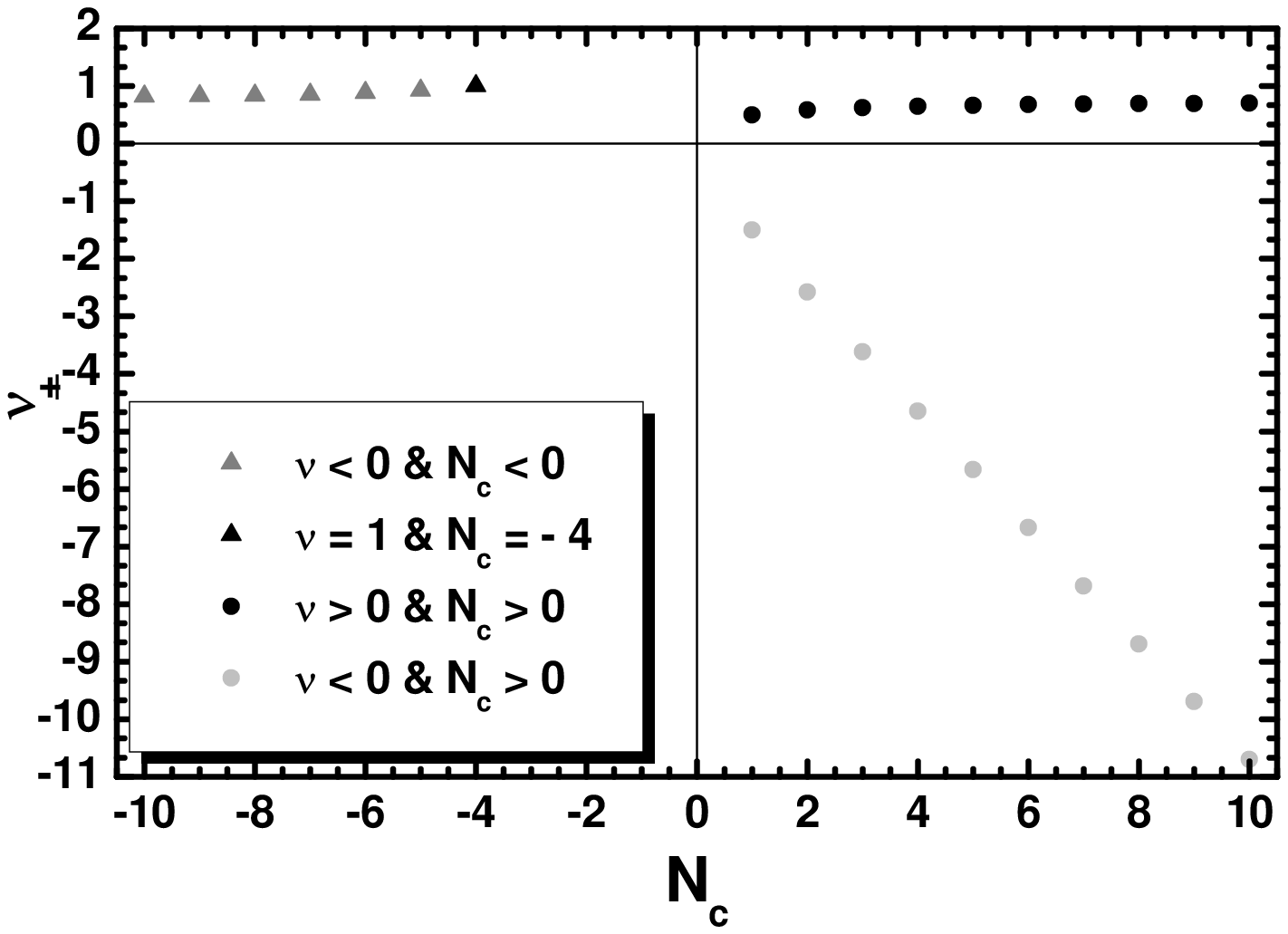}
\caption{\sl \small  Solutions to \Eref{nupm} for integer $\no$
values and $|\no|\leq10$. We take $\no<0$ (gray and black
triangles), or $\no>0$ and $\nu>0$ (black circles) or $\no>0$ and
$\nu<0$ (light gray circles).}\label{fignN}
\end{figure}
%%%%%%%%%%%%%%%%%%%%%%%%%%%%%%%%%%%%%%%%

\subsubsection{\small\sf Classification of the Solutions} \label{hd22}

To be more specific, we list in \Tref{tab1} some pairs $(\nu,\no)$
compatible with \eqs{sol}{all}. All of these correspond to
$\nu=\nm$ in \Eref{nupm} and yield integer $\no$ for $\nu=1$ and
$-3/2$. Confining ourselves on exclusively integer $\no$ --
favored by the string theories \cite{ibanez} --, we may vary $\no$
in the range $[-10, 10]$ and obtain $\nu$ via \Eref{nupm}. We
depict the results in \Fref{fignN}, where we present three
families of points of different shapes corresponding to different
types of $\Vhoc$ as we explain below. Besides the $\nu$ values
corresponding to $\no=-4$ and $1$, the others are irrational
numbers which, are used mainly for presentational purposes -- see,
e.g., \cref{noscale18}. Also, for $\no>0$ both outputs, $\np$ and
$\nm$, are possible and, in particular, the $\nm$ values increase
almost linearly with $\no$ whereas the $\np$ values remain almost
unchanged at the level of the values obtained for $\no<0$.

%In practice, \Eref{condn} is not sufficient for the achievement of
%Minkowski vacua since ...

\begin{figure}[!t]%\vspace*{-.25in}
\includegraphics[width=60mm,angle=-90]{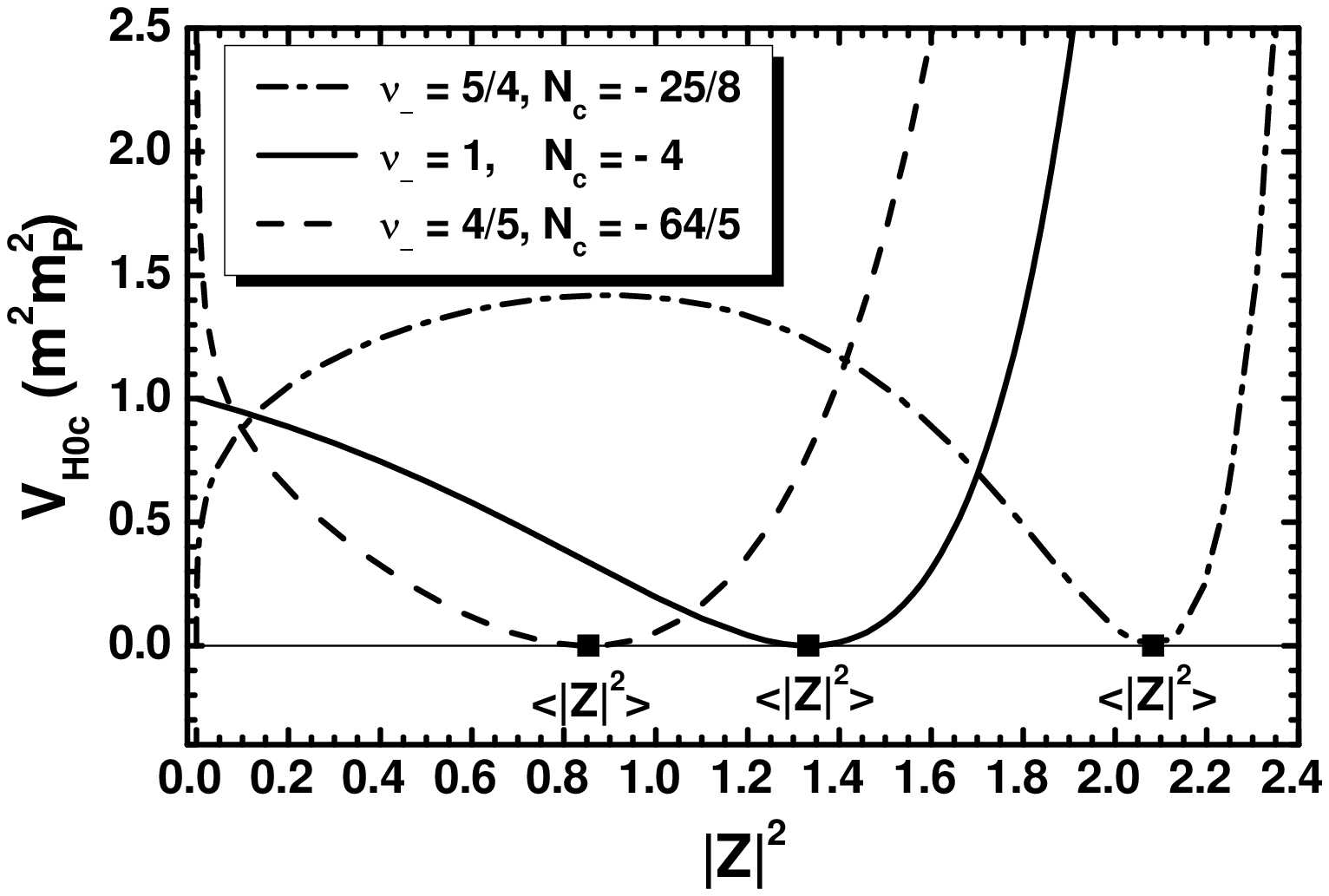}
\caption{\sl \small The (dimensionless) hidden-sector potential
$\Vhoc/m^2\mP^2$ in \Eref{Vh} as a function of $|Z|^2$ for
$(\nm,\no)=(4/5, -64/5)$, $(1, -4)$ and $(5/4, -25/8)$ (dashed,
solid and dot-dashed line respectively). The line $\Vhoc=0$ and
the values $\vev{|Z|^2}$ are also indicated.}\label{fig1}
\end{figure}

The structure of $\Vhoc$ in \Eref{Vh2} for the various pairs
$(\nu,\no)$ satisfying \Eref{sol} can be inferred by
Figs.~\ref{fig1} and \ref{fig2}, where we present $\Vhoc/m^2\mP^2$
as a function of $|Z|^2$ for $\no<0$ -- in Fig.~\ref{fig1} -- and
for $\no>0$ -- in Fig.~\ref{fig2}. Namely, we draw solid, dashed
and dot-dashed lines for $(\nm,\no)=(1, -4)$, $(4/5, -64/5)$ and
$(5/4, -25/8)$ respectively in \Fref{fig1}. On the other hand, the
dashed, solid and dot-dashed lines in \Fref{fig2} are obtained for
$(\np,\no)=(1/2, 1)$, $(\nm,\no)=(-3/2, 1)$ and $(\nm, \no)=(-2,
16/11)$ respectively. The line $\Vhoc=0$ and the values
$\vev{|Z|^2}$ at the non-SUSY minimum -- see \Eref{vev} -- are
also indicated in both figures. More explicitly, the specific
values of $\vev{|Z|^2}$ for each $\nu$ value employed in the
figures above are arranged in \Tref{tab2}. We see that
$\vev{|Z|^2}$ decreases with $|\nu|$ and may become even
subplanckian, in sharp contrast to the well-known Polonyi model
\cite{polonyi} and the case of \cref{susyr}.

%%%%%%%%%%%%%%%%%%%%%%%%%%%%%%%%%%%%%%%%%%%%%%%%%%%%%%%%%%%%%%%%%%%%
%%%%%%%%%%%%%%%%%%%%%%%%%%%%%%%%%%%%%%%%

 %-- according to \Eref{vev}.

\begin{table}[b] \caption{\normalfont Minkowski vacua  for
the $\nu$ values used in Figs.~\ref{fig1} and \ref{fig2}.}
\begin{ruledtabular}
\begin{tabular}{c||ccc|ccc}
&\multicolumn{3}{c|}{\Fref{fig1}}&\multicolumn{3}{c}{\Fref{fig2}}\\\hline
$\nu$&  $5/4$&$1$ & $4/5$&$-2$&$-3/2$& $1/2$\\
$\vev{|Z|^2}$&  $25/12$&$4/3$ & $64/75$&$16/3$& $3$&$1/3$
\end{tabular}
\end{ruledtabular} \label{tab2}
\end{table}
%%%%%%%%%%%%%%%%%%%%%%%%%%%%%%%%%%%%%%%%

Let's focus first on \Fref{fig1}. We remark that for $\nu>1$ in
the domain of \Eref{all}, $\Vhoc$ features a SUSY minimum at
$\vev{|Z|}=0$ apart from the non-SUSY Minkowski one -- see
dot-dashed line in \Fref{fig1}. This is due to the presence of the
factor $|Z|^{2(\nu-1)}$ in \Eref{Vh} which yields another solution
to the right equation in \Eref{vev} for $\nu>1$. Although less
attractive than the other cases, the coexistence of SUSY and
non-SUSY Minkowski vacua may be beneficial for applying the
multi-point principle analyzed in \cref{nevz}. We obtain the same
shape of $\Vhoc$ for $1<\nu<3/2$. Since no integer $\no$ value may
be computed via \Eref{sol} for $\nu$ values into this domain, no
point in \Fref{fignN} corresponds to this kind of $\Vhoc$. For the
value $(\nu, \no)=(1,-4)$ -- black triangle in \Fref{fignN} -- we
obtain the well-known form of $\Vhoc$ presented in \cref{susyr},
whereas for the remaining (gray) triangles in \Fref{fignN}  we
obtain the type of $\Vhoc$ drawn by dashed line in \Fref{fig1}. It
is remarkable that $\Vhoc$ develops just one critical point in
this case.  The situation is rather different for $\no>0$ as shown
in \Fref{fig2}. For $\nu>0$ -- e.g., see the $(\nu, \no)$ values
represented by black circles in \Fref{fignN} --, the type of
$\Vhoc$ corresponds to the dashed line in \Fref{fig2}. On the
contrary, the $\nu<0$ values give rise to $\Vhoc$ whose structure
is designed by the solid and the dot-dashed lines. As $\no$
increases, $|\no|$ increases too -- see e.g., the gray circles in
\Fref{fignN} -- and $\vev{|Z|^2}$ is moved towards larger values.

\subsubsection{\small\sf Particle Spectrum} \label{hd23}

If we analyze $Z$ according to the description \beq
Z=(z+i\th)/\sqrt{2}\label{Zpara}\eeq and expand $\Vhoc$ in
\Eref{Vh2} about the configuration
\beq\vev{z}=2\sqrt{\frac23}|\nu|~~~\mbox{and}~~\vev{\th}=0,\label{z0}\eeq
-- cf. \Eref{vev} --, we can work out the hidden-sector spectrum
of the model. This is composed of:

\begin{itemize}

%%%%%%%%%%%%%%%%%%%%%%%%%%%%%%%%%%%%%%%%%%%%%%%%%%%%%%%%%%%%%%%%%%%%
\begin{figure}[!t]%\vspace*{-.25in}
\includegraphics[width=60mm,angle=-90]{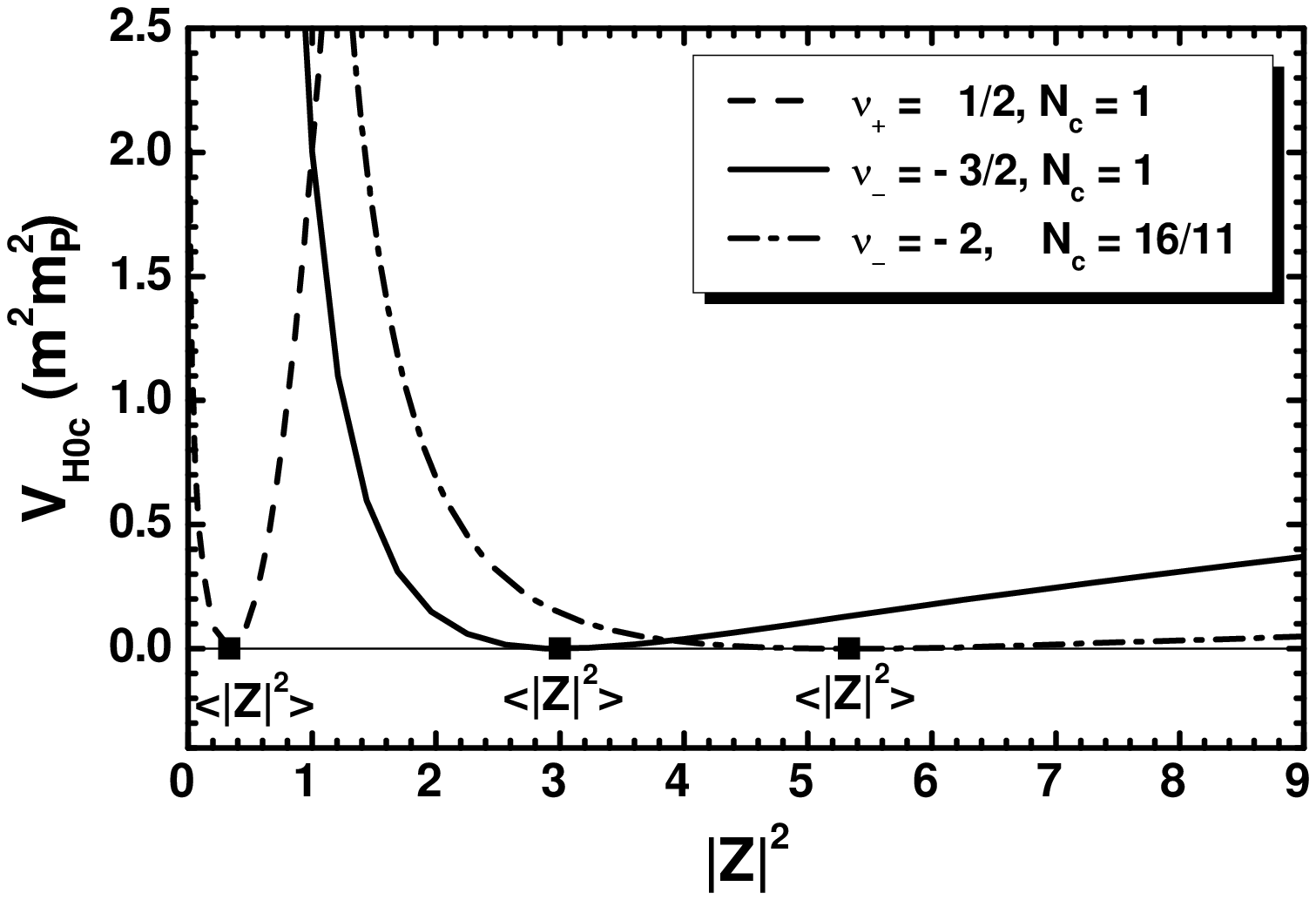}
\caption{\sl \small The same as in \Fref{fig1} but for
$(\nu,\no)=(1/2,1)$, $(-3/2, 1)$, and $(-2, 16/11)$ (dashed, solid
and dot-dashed line respectively).}\label{fig2}
\end{figure}

\item[{\ftn\sf (i)}] A (canonically normalized) real scalar field
-- the sgolstino or $R$ saxion --, $\what z=\sqrt{\vev{\rm g}}z$,
with mass squared
\beqs\beq m^2_{\what z} =\vev{\partial^2_{\what
z}\Vhoc}=\vev{{2\Vhoc''|Z|^2}/{{\rm
g}}}=\frac94\frac{\mgro^2}{|\nu|^{2}\vevg}; \label{mz0}\eeq
%|\nu|^{-2}\vevg^{-1}\mgro^2
where we take into account \eqs{Vhp}{defg},

\item[{\ftn\sf (ii)}] The gravitino, $\Gr$,  -- which absorbs the
fermionic partner of the $R$ saxion, the $R$ axino -- with mass
squared
\beq \label{mgr0} \mgro^2=m^2\vev{e^{\khc}|Z|^{2\nu}}=4^{\nu}
|\nu|^{2\nu}m^2/3^\nu\vevg^{\no/2}.\eeq\eeqs
%\,\frac{}{\vevg^{\no/2}}

\item[{\ftn\sf (iii)}] A massless Nambu-Goldstone boson, $\th$, --
referred \cite{rnelson} to as an $R$ axion.

\end{itemize}

Besides the last one, the masses above keep their values even for
$k\neq0$ and obey the super-trace formula \cite{nilles} which, in
the case of an hidden sector with one superfield, reads
\cite{susyr}
\beq \label{tr0} {\sf
STr}M_0^2=\mzo^2-4\mgro^2=6\mgro^2\vev{{\mathcal R}^{(0)}_{\rm
Hc}}=\frac{12}{\no}\mgro^2,\eeq
where we make use of Eqs.~(\ref{ds0}), (\ref{sol}), (\ref{mgr0})
and (\ref{mz0}).

\subsection{\sc\small\sffamily  Including the $R$-Symmetry-Violating Term}
\label{hd3}

The (canonically normalized) $R$ axion, $\what\theta$, couples to
the massless SM gauge bosons (gluons and photons) \cite{ribe,
ribe1} due to $SU(3)_{\rm c}$ and $U(1)_{\rm EM}$ anomalies
respectively. Non-perturbative $SU(3)_{\rm c}$ instanton effects
\cite{anomalies} result in a mass for $\what\theta$ which is
expected to be at the KeV level and so it is severely constrained
by astrophysical and laboratory probes. Here, we focus on the most
robust lower bound derived from the fact that $\what\theta$ can be
produced in the supernovae core via the bremsstrahlung process
$nn\to nn\what\theta$ with $n$ being nucleons. The measurements of
the neutrino burst from supernova 1987a set important constraints
on sub-GeV axion-like particles. In particular, the energy loss
rate due those particle should not exceed the one through
neutrinos. Considering that the coupling between $\what\theta$ and
gluons or photons is well suppressed, the restriction above is
translated to the following bound on the mass of $\what\theta$
\cite{astro} \beq\mth\geq10~{\rm MeV}, \label{snv}\eeq where we
restore the units for convenience. On the other hand, we do not
impose here an upper limit on $\mth$ from a possible candidacy of
$\what\theta$ for cold dark matter since this depends crucially on
the cosmological evolution of $\what\theta$ which is not specified
in our set-up \cite{ribe}.

The phenomenological problems arising from the masslessness (at
perturbative level) of $\what\theta$ for $k=0$ in \Eref{khi} may
be evaded invoking several modifications of the setting in
\Sref{hd2}. E.g., if we consider $Z$ as a nilpotent superfield
\cite{nilpotent} no sgoldstino appears in the particle spectrum
and so no $R$ axion too. Another way out of this problem is the
gauging of the $R$ symmetry \cite{gaugedR, Rgauged, castano},
provided that it is anomaly-free. In this case, though, D-term
contributions are introduced which destabilize the construction of
\Sref{hd22}. Following \cref{susyr}, we here adopt the simplest
solution: we explicitly break the $R$ symmetry via a subdominant
quartic term in $\khi$ proportional to $k$ -- see \Eref{khi}. As
we show below, even for $k\neq0$, the field configuration in
\Eref{z0} still corresponds to a Minkowski vacuum whereas a large
enough mass for the $R$ axion is generated. This is actually the
reason for which we prefer to activate the $R$-violation in $\khi$
and not in $\whi$ as in the Polonyi model \cite{polonyi}. It is
clear that no purely theoretical motivation exists for this term
in $\khi$. Its presence, though, is ``technically natural'' thanks
to the smallness of $k$ -- see below. Although not generalized
here, the exponent, $4$, of this term is the unique which offers
the ``facilities'' above \cite{susyr}. The same term is widely
utilized for the stabilization of the imaginary direction of the
sgoldstino within the no-scale models \cite{noscale18}.

Below we verify that $(\vev{z}, \vev{\th})$ defined in \Eref{z0}
represents a stable Minkowski vacuum of $\Vhi$ in \Eref{Vhzz} for
$k\neq0$ and $N=\no$. To embark on it, we express $\Vhi$ as a
function of $z$ and $\th$ using the parametrization in
\Eref{Zpara}. First of all, we show that $\Vhc$ at the point given
by \Eref{z0} vanishes, i.e.,
\beq
\vev{\Vhc}=0~~\Rightarrow~~\vev{\uc\vc/\wc}=3\no\vev{|Z|^2}.\label{Vhvev}\eeq
Indeed, \eqss{u}{v}{w} at the vacuum yield \beq \label{uvw0}
\vev{\uc}=\vev{\vc}=\nu\no+(\no+\nu)\vev{|Z|^2}~~\mbox{and}~~\vev{\wc}=\no,\eeq
and so \Eref{Vhvev} can be readily deduced. Then we minimize
$\Vhc$ w.r.t $z$ and $\theta$. Given that $\zm=\sqrt{2}i\th$,
$\Vhc$ for $\th=0$ coincides with the one in \Eref{Vh2} , i.e.,
\beq \Vhc(z,\th=0)=\Vhoc\label{Vhz}\eeq
and therefore, $\vev{z}$ (for $k\neq0$) keeps its value in
\Eref{z0} (found for $k=0$). As regards the $\th$ direction, we
check the validity of the extremum condition computing the first
derivative of $\Vhc$ w.r.t $\th$ for $\vev{\th}=0$ with result
\begin{align}\nonumber \vev{\partial_{\th}\Vhc}&= {3m^2}\vev{e^{\khc}
|Z|^{2(\nu-1)}}\cdot\\&\vev{\lf\frac{\partial_{\th}\uc}{\uc}+\frac{\partial_{\th}\vc}{\vc}-\frac{\partial_{\th}\wc}{\wc}\rg|Z|^2-\th}
=0.\label{vt3}
\end{align}
Here we take advantage of \Eref{Vhvev}. We also observe that
\beqs\beq\vev{\partial_{\th}\lf e^{\khi} |Z|^{2(\nu-1)}\rg}=0\eeq
since this is proportional at least to $\th^2$ and, similarly,\beq
\vev{\partial_{\th}\wc}=\vev{\partial_{\th}\uc}=\vev{\partial_{\th}\vc}=0,\eeq\eeqs
since these are proportional to $\th$ -- see \eqss{w}{u}{v}.

From the results above, it is easy to compute the mixed second
derivatives
\beq \vev{\partial_{z}\partial_{\th}\Vhc}=
\vev{\partial_{\th}\partial_{z}\Vhc}=0. \eeq
which are also zeroed. Therefore, the matrix of the second
derivatives of $\Vhi$ w.r.t $z$ and $\th$,
$\vev{\partial_{i}\partial_{j}\Vhc}$ with $i=z$ and $\th$ has a
diagonal form. For $i=j=z$ we obtain the mass squared of $\what
z$, given by \Eref{mz0}, since the $z$-dependent form of $\Vhi$ in
\Eref{Vhzz} coincides with that of $\Vhoc$ in \Eref{Vh2}. For
$i=j=\th$, we differentiate once more $\wc, \uc$ and $\vc$ w.r.t
$\th$ with results \beqs\begin{align}
\vev{\partial^2_{\th}\wc}&=-48k\lf\vev{|Z|^2}+\no\rg,\label{aux1}\\
\vev{\partial^2_{\th}\uc}&=\vev{\partial^2_{\th}\vc}=(\nu+\no).\label{aux2}\end{align}\eeqs
Armed with the results above we arrive at the following one,
\begin{align}\vev{\partial_{\th}^2\Vhc}&=
{3m^2}\vev{e^{\khc}
|Z|^{2(\nu-1)}}\cdot\nonumber\\&\Bigg(\vev{|Z|^2}\vev{\frac{\partial^2_{\th}\uc}{\uc}+\frac{\partial^2_{\th}\vc}{\vc}-\frac{\partial^2_{\th}\wc}{\wc}}-1\Bigg)\label{vt4}
\end{align}
Making use of the identity \beq
\label{aux}\vev{|Z|^2}\lf\frac{\partial^2_{\th}\uc}{\uc}+\frac{\partial^2_{\th}\vc}{\vc}\rg=1\eeq
and canonically normalizing the relevant mode, we may achieve our
final result
\beq\mth^2=\vev{\partial^2_{\what \th}\Vhc}
=\vev{\partial^2_{\th}\Vhc/{\rm g}}= 144k\mgr^2/\vevg^{3/2},
\label{mth}\eeq
which coincides with the one obtained in \cref{susyr} for $\nu=1$.
We observe that $\mth^2>0$ for $k>0$. Consequently, setting $k>0$
in \Eref{khi} does not modify $\Vhi$ from $\Vhoc$ in the real
direction but just allows for a non-vanishing $R$-axion mass. To
fulfill \Eref{snv} for, e.g., $m=1~\TeV$, it is enough to take
$k\gtrsim4\cdot10^{-13}$ -- this number corrects a typo in
\cref{susyr} -- for $\nu=\np>0$ and $k\gtrsim2\cdot 10^{-14} -
3\cdot10^{-5}$ for $\nu=\nm<0$, where the minimal $k$ increases as
$\no$ increases from $1$ to $10$ -- see \Fref{fig1}. Therefore,
our argument related to the naturalness of the $k$ term in
\Eref{khi} is compatible with the $R$-axion phenomenology.

Summarizing, the particle spectrum of the present version of our
model comprises  $\Gr$, $\what z$ and $\what \th$ whose the masses
are given by \eqss{mgr0}{mz0}{mth}, respectively. We can verify
that these masses obey the supertrace relation -- cf. \Eref{tr0}
\bea\nonumber {\sf STr}M^2
&=&\mz^2+\mth^2-4\mgr^2=6\mgr\vev{{\mathcal R}_{\rm
Hc}}\\&=&\frac{9}{\nu}\lf\frac1\nu-\frac43\lf1-12k\nu/\vevg^{3/2}\rg\rg\mgr^2,
\label{trkl}\eea
where $\vev{{\mathcal R}_{\rm  Hc}}$ is geometrically \cite{susyr}
estimated as
\beq\label{RHk}  \vev{{\mathcal R}_{\rm  Hc}}=\vev{{\mathcal
R}^{(0)}_{\rm Hc}}-24k/\vevg^{3/2},\eeq
with $\vev{{\mathcal R}^{(0)}_{\rm Hc}}$ given in \Eref{ds0}.
Contrary to the case with $k=0$, here ${\mathcal R}_{\rm  Hc}$ is
not generically constant.

Checking the hierarchy of the various masses, we infer from
\eqs{mgr0}{mz0} that \beqs \beq
\mgr\leq\mz~~\mbox{for}~~\nu\vev{\rm g}^{1/2}\leq3/2
\label{mzgr}\eeq and from \eqs{mgr0}{mth} that \beq
\mth\leq\mgr~~\mbox{for}~~k\leq\vev{\rm
g}^{3/2}/144.\label{m8gr}\eeq \eeqs More specifically, we obtain
$\mz\leq2\mgr$ for $\nu>0$ and $\no<0$ and so the decay of $\what
z$ into a pair of $\Gr$ \cite{ribe} is kinematically blocked
whereas for $\no>0$ this decay channel is open. Therefore, the
decay of $\what z$ and $\what \th$ into $\Gr$ depends on the
parameters ($\nu$, $\no$ and $k$) and may have an impact on the
relic abundance of $\Gr$ before nucleosynthesis.

%%%%%%%%%%%%%%%%%%%%%%%%%%%%%%%%%%%%%%%%%%%%%%%%%%%%%%%%%%%%%%%%%%%%
\begin{figure}[!t]%\vspace*{-.25in}
\includegraphics[width=60mm,angle=-90]{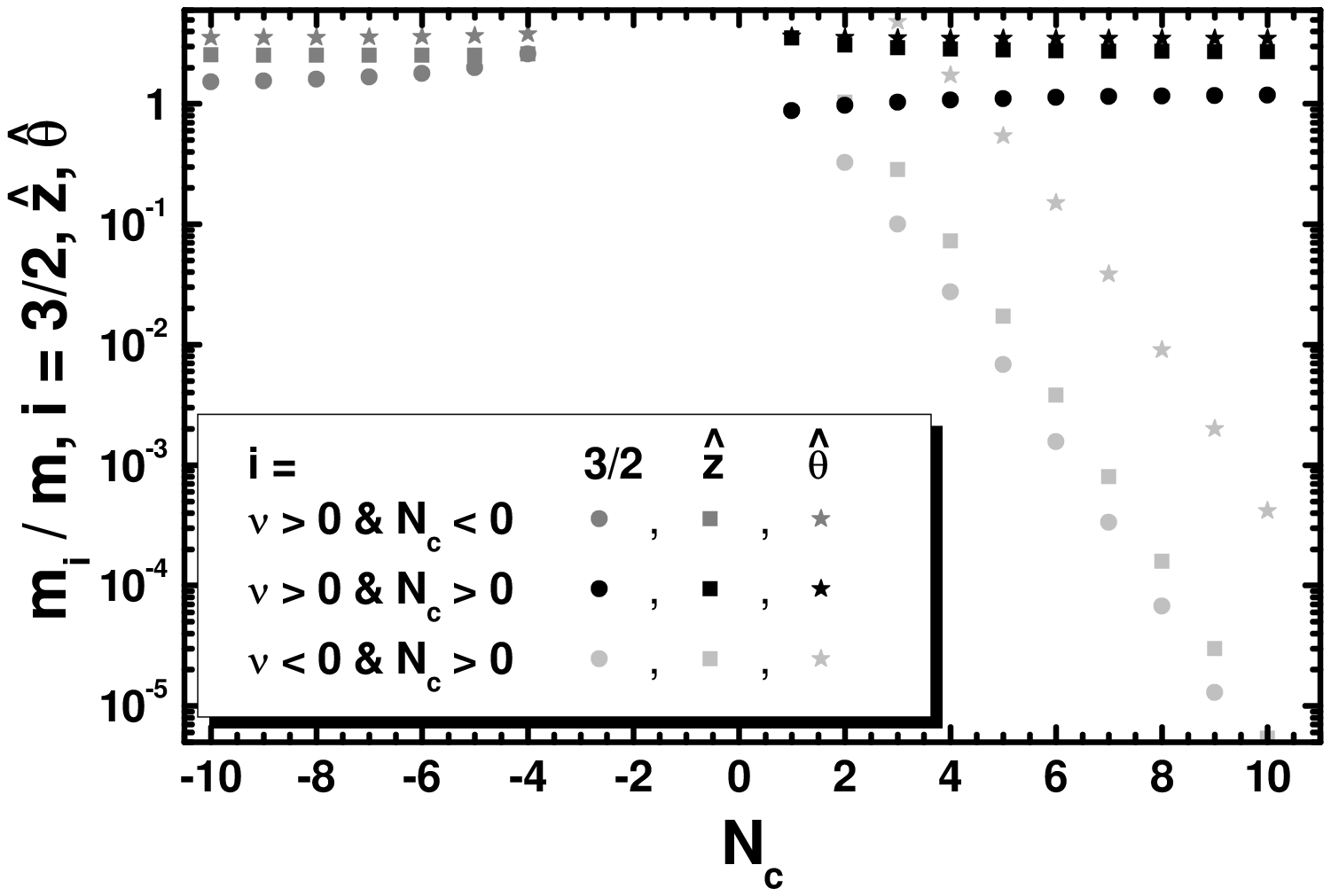}
\caption{\sl \small The ratios $\mgr/m$, $\mz/m$ and $\mth/m$ for
$k=0.05$ (circles, squares and stars respectively) versus $\no$.
We take $\no<0$ (gray symbols) or $\no>0$ and $\nu>0$ (black
symbols) or $\no>0$ and $\nu<0$ (light gray symbols).}\label{fig5}
\end{figure}
%%%%%%%%%%%%%%%%%%%%%%%%%%%%%%%%%%%%%%%%

These conclusions are illustrated in \Fref{fig5}, where we display
the ratios $\mgr/m$ (circles), $\mz/m$ (squares) and $\mth/m$
(stars) for $k=0.05$ versus $\no$, for integer $\no$ values and
$|\no|\leq10$. We take $\no<0$ (gray symbols) or $\no>0$ and
$\nu>0$ (black symbols) or $\no>0$ and $\nu<0$ (light gray
symbols). For the selected $k$ value $\what \th$ is heavier than
both $\what z$ and $\Gr$ for any $(\nu,\no)$ whereas the hierarchy
of $\what z$ and $\Gr$ is $k$-independent and mildly dependents on
$(\nu,\no)$ according to \Eref{mzgr}.

Just for completeness, we mention that the $R$ axion may acquire
mass squared
\beq \mth^2= 144k\mgr^2/\vevg^{2} \label{mtha}\eeq similar to the
one in \Eref{mth}, if we adopt one of the following \Ka s
\beqs\bel K_{\rm H} &=\no\ln\lf1+|Z|^2/\no\rg+N_k\ln\lf1-k\zm^4/N_k\rg,\label{kh1} \\
K_{\rm H} &=\no\ln\lf1+|Z|^2/\no\rg-k\zm^4,
\label{kh2}\end{align}\eeqs
where $N_k$ is an undetermined constant.

\section{Observable Sector}\label{obs}

Our next task is to study the transmission of the SUSY breaking to
the visible world. To implement this, we introduce the chiral
superfields of the observable sector $\phc_\al$ and assume the
following structure -- cf.~\cref{polonyi,susyr} -- for the total
superpotential, $W$, and \Ka, $K$, of the theory
\beqs\bea \label{Who} W&=\whi(Z) + W_{\rm O}\lf\phc_\al\rg,\\
\label{Kho} K&=\khc(Z)+\wtilde K(Z)|\phc_\al|^2,\eea\eeqs
where $\whi$ and $\khc$ are given by \eqs{whi}{khi} with $N=\no$
whereas $W_{\rm O}$ and $\wkhi$ are specified in \Sref{obs1} for a
generic SUSY model and, in \Sref{obs2}, for the MSSM. In the
latter case, a solution to the $\mu$ problem is also proposed.

\subsection{\sc\small\sffamily Generic Model}\label{obs1}

Here we adopt the following quite generic form of $W_{\rm O}$
\beq W_{\rm O}=h \phc_1\phc_2\phc_3+\mu\phc_4\phc_5,
\label{w0}\eeq
where we may easily select the appropriate  $R$ charge for each of
$\phc_\al$ as in \cref{susyr}. We try also similar $\wkhi$,
ensuring universal SSB parameters for $\phc_\al$, i.e.,
\beqs\bel
K_1&=\khc+\mbox{$\sum_\al$}|\phc_\al|^2,~~~\label{K1}\\
K_2&=\no\ln\Big(1+\lf|Z|^2-k\zm^4+\mbox{$\sum_\al$}|\phc_\al|^2\rg/\no\Big),\label{K2}\\
K_3&=\khc+N_{\rm O}\ln\left(1+\mbox{$\sum_\al$}|\phc_\al|^2/N_{\rm
O}\right),\label{K3} \end{align}\eeqs
where the specific value of $N_{\rm O}$ is irrelevant for our
purposes. If we expand the $K$'s above for low $\phc_\al$ values,
these may assume the form of \Eref{Kho} with $\wkhi$ being
identified as
\beq \label{wk} \wkhi=\begin{cases} 1&\mbox{for}~~K=K_1,
K_3;\\\lf1+(|Z|^2-k\zm^4)/\no\rg^{-1}&\mbox{for}~~K=K_2\,.\end{cases}\eeq
Compared to the $K$'s adopted in \cref{susyr}, we remark that the
denominator $\no$ in the equations above is replaced by $-4$
there. This is obvious since $\nu=1$ in \Eref{whi} is associated
with $\no=-4$ in \Eref{khi}.

Adapting the general formulae of \cref{soft} to the case with one
hidden-sector field, as done in \cref{susyr}, we obtain the SSB
terms in the effective low energy potential which can be written
as
\beq V_{\rm SSB}= \wtilde m_\al^2 |\what\phc_\al|^2+\lf Ah
\what\phc_1\what\phc_2\what\phc_3+ B\mu
\what\phc_4\what\phc_5+{\rm h.c.}\rg, \label{vsoft} \eeq
where the canonically normalized fields
$\what\phc_\al=\veva{\wtilde K}^{1/2}\phc_\al$ are denoted by
hats. In deriving the values of the SSB parameters above, we find
it convenient to distinguish the cases:

\subparagraph{\sl (a)} For $K=K_1$ and $K_3$, we see from
\Eref{wk} that $\wkhi$ is constant and so the relevant derivatives
are eliminated. Substituting
\beqs\bel\vev{F^Z}&=\veva{\bar
F^\bz}=\sqrt{3}\mgro/\vevg^{1/2},\\~\vev{e^{\khc}}&=\vevg^{-\no/2},
~\vev{\partial_Z\khc}=\frac{2\nu}{\sqrt{3}}\vevg^{1/2}\label{aux1f}\end{align}\eeqs
into the relevant expressions \cite{susyr} we arrive at
%
%\bea
\beq \wtilde
m_\al=\mgro~~~\mbox{and}~~~A=B+\mgro=2\nu\mgro/\vevg^{\no/4}\,.\label{mk1}\eeq

\subparagraph{\sl (b)} For $K=K_2$, $\wkhi$ in \Eref{wk} is $Z$
dependent with $\veva{\wkhi}=\vevg^{1/2}$ and the relevant
derivatives are found to be
\beqs\bel
\vev{\partial_Z\ln\wkhi^2}&=\frac{2}{3}\vev{\partial_Z\ln\wkhi^3}=-\frac{4\nu}{\sqrt{3}\no}\vevg^{1/2},\\
\vev{\partial_\bz\partial_Z\ln\wkhi}&=\frac{4\nu^2}{3\no^2}\vevg-\frac{1}{\no}\vevg^{1/2}.\label{aux2f}\end{align}\eeqs
Inserting the expressions above into the general formulae
\cite{susyr} we end up with
\beqs\bel \label{mk2m}\wtilde m_\al &=\frac{3}{4|\nu|}\mgro/\vevg^{1/2},\\
\label{mk2A} A &=\frac{9}{8\nu}\mgro/\vevg^{(\no+7)/4},\\ B&=
\frac{3}{2\nu}(1-\nu)\mgro/\vevg^{(\no+4)/4},
\label{mk2}\end{align}\eeqs
%
%\eea
where the last result reveals the reason for which $B$ vanishes
for $\nu=1$ \cite{susyr}.

%%%%%%%%%%%%%%%%%%%%%%%%%%%%%%%%%%%%%%%%%%%%%%%%%%%%%%%%%%%%%%%%%%%%
\begin{figure}[!t]%\vspace*{-.25in}
\includegraphics[width=60mm,angle=-90]{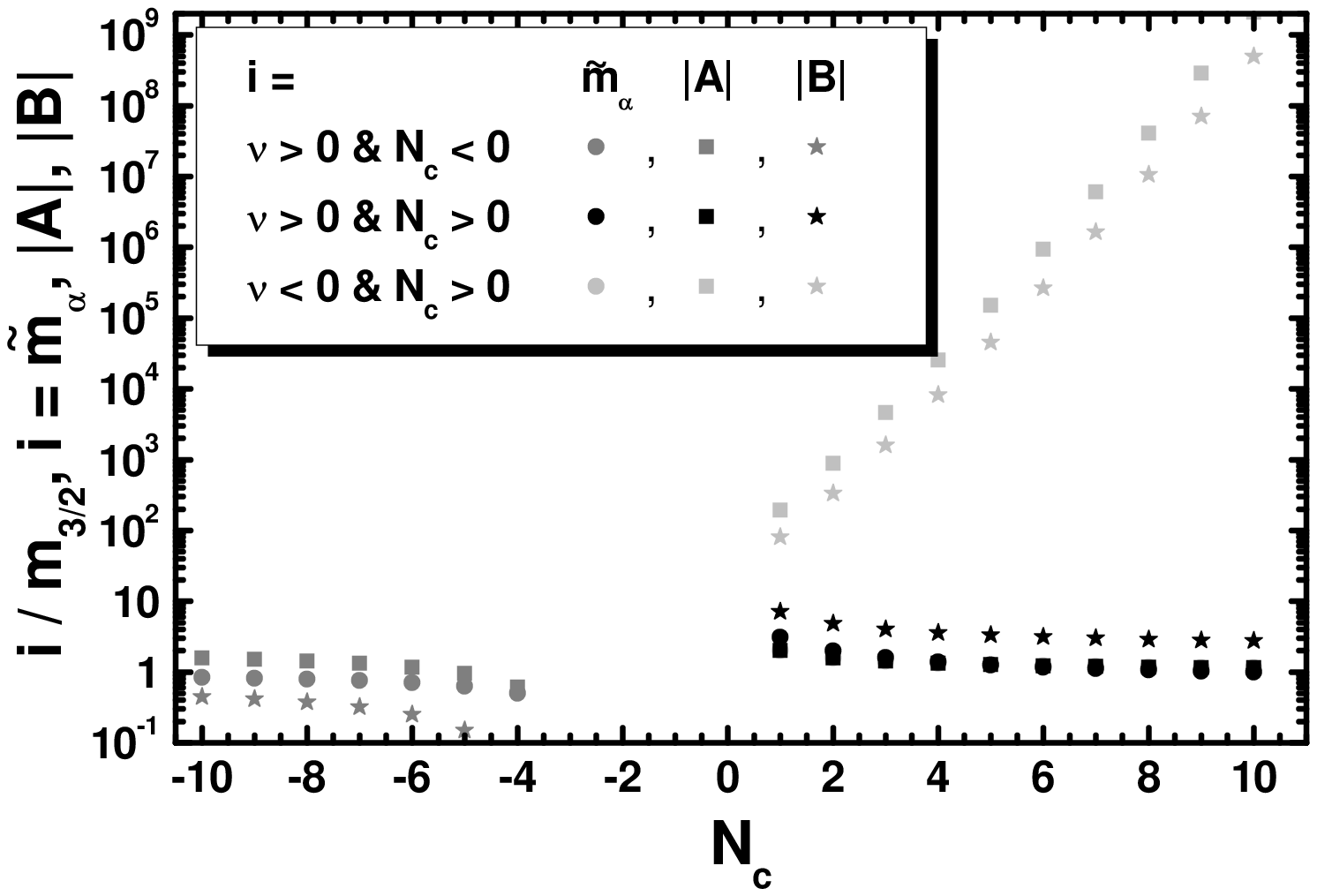}
\caption{\sl \small The ratios $m_\al/\mgr$, $A/\mgr$ and $B/\mgr$
for $K=K_2$ (circles, squares and stars respectively) versus
$\no$. We take $\no<0$ (gray symbols) or $\no>0$ and $\nu>0$
(black symbols) or $\no>0$ and $\nu<0$ (light gray
symbols).}\label{fig6}
\end{figure}
%%%%%%%%%%%%%%%%%%%%%%%%%%%%%%%%%%%%%%%%

The hierarchy of the SSB parameters above w.r.t $\mgr$ is
demonstrated in \Fref{fig6}, where we display the ratios $\wtilde
m_\al/\mgr$ (circles), $|A|/\mgr$ (squares) and $|B|/\mgr$ (stars)
versus $\no$, for integer $\no$ values and $|\no|\leq10$. We take
$\no<0$ (gray symbols) or $\no>0$ and $\nu>0$ (black symbols) or
$\no>0$ and $\nu<0$ (light gray symbols). Note that the $\wtilde
m_\al$'s corresponding to the same $\no>0$ turn out to be equal
for $\nu<0$ and $\nu>0$ since these are inverse proportional to
$\vev{\rm g}^{1/2}\nu$ which is constant for the same $\no$, as
deduced from the last relation in \Eref{defg}. For this reason the
black circles overlap the light grays ones in the graph. We
observe that in all cases the $\wtilde m_\al$'s are of the order
of $\mgr$ whereas $|A|$ and $|B|$ may be much larger than it for
$\nu<0$ and $\no>0$. This is due to the fact that the $\nu=\nm$
values, for these $\no$'s, may be much larger than unity -- see
light gray circles in \Fref{fig1}. Therefore, we expect that the
produced sparicle spectrum will be rather heavier than $\mgr$ in
these cases. Similar conclusions can be extracted for $K=K_1$ too.

\subparagraph{} Let us emphasize, finally, that the presence of
$k\neq0$ in \Eref{khi} not only provides mass to the $R$ axion but
also breaks explicitly $U(1)_R$ and so, no topological defects are
generated when $Z$ acquires its v.e.v in \Eref{z0}. Otherwise,
$\vev{z}$ could break the discrete subgroup to which $U(1)_R$ is
broken, due the SSB parameters in $V_{\rm SSB}$, and the
production of topological defects could be possible.

\subsection{\sc\small\sffamily Generation of the $\mu$ Term of MSSM}
\label{obs2}

Extending further the results in \cref{susyr}, we now check if the
hidden-sector models introduced in this work offer an explanation
of the $\mu$ term of MSSM, following the recipe of Guidice and
Masiero in \cref{masiero}. To this end, we assign the $R$ charges
for the MSSM fields indicated in \cref{susyr}, which forbid terms
of the form $\hu\hd$ in $W_{\rm O}$ -- see \Eref{w0}. Namely, we
may write it as
\beq W_{\rm MSSM} =h_{\al\bt\gamma}
\phc_\al\phc_\bt\phc_\gamma/6\,,\label{wmssm}\eeq
where we closely follow the notation established in \cref{susyr}.
The mixing term between $\hu$ and $\hd$ may emerge in the part of
the potential including the SSB terms
\bea \nonumber V_{\rm SSB}&=& \wtilde m_\al^2
|\what\phc_\al|^2+\lf\frac16 A_{\al\bt\gamma} h_{\al\bt\gamma}
\what\phc_\al\what\phc_\bt\what\phc_\gamma\right. \\&+& \wtilde
B\mu \what H_u\what H_d+{\rm h.c.}\Big), \label{vmssm} \eea
if we incorporate (somehow) into the $K$'s of \Erefs{K1}{K3} the
following higher order terms, inspired by \cref{masiero},
\beq \dK=\lm\frac{\bz^{2\nu}}{\mP^{2\nu}}\hu\hd\ +\ {\rm
h.c.},\label{dK}\eeq
where $\lm$ is a real constant and the hatted fields in
\Eref{vmssm} are related to the unhatted ones as shown below
\Eref{vsoft}.  Due to the adopted $R$ symmetry, the terms in
\Eref{dK} do not coincide with those proposed in the original
paper \cite{masiero}. It is, therefore, non-trivial to find out
the magnitude of the resulting $\wtilde B\mu$.

We consider several \Ka s to exemplify our approach, as done in
\cref{susyr}. Namely, we select
\beqs\bel
K_{11}&=K_1+\dK,~~~\label{K11}\\
K_{21}&=K_2+\dK,\label{K21}\\
K_{22}&=\no\ln\left(1+\left(|Z|^2-k\zm^4+\mbox{$\sum_\al$}|\phc_\al|^2+\dK\right)/\no\right),~~~~~~\label{K22}\\
K_{23}&=\no\ln\Big(1+\left(|Z|^2-k\zm^4+\dK\right)/\no\Big)+\mbox{$\sum_\al$}|\phc_\al|^2,~~~~~~~~\label{K23}\end{align}\eeqs
\\ [-0.4cm]
where $K_1$ and $K_2$ are defined in \eqs{K1}{K2} respectively.
The $K$'s above may be brought into the form
\beq K_{\rm MSSM}=\khc(Z)+\wtilde
K(Z)|\phc_\al|^2+\big(\km\hu\hd+{\rm h.c.}\big), \label{kmssm}\eeq
where $\wkhi$ is determined as shown in \Eref{wk}. Namely the
branch of the definition of $\wkhi$ in \Eref{wk} for $K=K_1$ and
$K_3$ corresponds to $K=K_{11}$ and $K_{23}$, whereas that one
which is valid for $K=K_2$ corresponds to $K=K_{21}$ and $K_{22}$.
As a consequence, the derived $\wtilde m_\al$ and
$A_{\al\bt\gamma}$ in \Eref{mk1} remain the same for $K=K_{11}$
and $K_{23}$, whereas those shown in \eqs{mk2m}{mk2A} are found
also for $K=K_{21}$ and $K_{22}$.

For the computation of $\wtilde B\mu$ in \Eref{vmssm} it is
crucial to find out $\km$ in \Eref{kmssm}. This is done by
expanding the $K$'s in \Erefs{K11}{K23} for low $\hu$ and $\hd$
values. Our result is \beq \label{km}
\km=\lm\frac{\bz^{2\nu}}{\mP^{2\nu}}\begin{cases}
1&\mbox{for}~~K=K_{11},
K_{21};\\
\Big(1+\frac{|Z|^2-k\zm^4}{\no}\Big)^{-1}&\mbox{for}~~K=K_{22},K_{23}.\end{cases}
\eeq
Thanks to the non-vanishing $\km$, we expect that the effective
coefficient $\wtilde B\mu$ in \Eref{vmssm} assumes a non-vanishing
value which may be found by applying the relevant formula in
\cref{soft,susyr}. In particular, we obtain
\beq \label{Bm} \frac{\wtilde B\mu}{\mgr^2}=
\lm\lf\frac{4\nu^2}{3}\rg^\nu\cdot\begin{cases}4(\nu-1)&\mbox{for}~~K=K_{11};
\\3(6\nu-5)/2\nu\vevg&\mbox{for}~~K=K_{21};\\ 9(1-2\nu)/8\nu^2\vevg&\mbox{for}~~K=K_{22};
\\ 3(2-3\nu)/\nu&\mbox{for}~~K=K_{23},
\end{cases}\eeq
where we take into account the following:

\subparagraph{\sl (a)} For $K=K_{11}$ and $K_{21}$, $\km$ in
\Eref{km} is only $\bz$ dependent and so we have
\beqs\beq
\vev{\km}=\vev{\partial_\bz\km}/\sqrt{3}=\lm\lf4\nu^2/3\rg^\nu\,.\label{aux3}\eeq
As regards $\wkhi$, this is trivial for $K=K_{11}$, whereas for
$K=K_{21}$, its derivatives w.r.t $Z$ can be computed with the aid
of \Eref{aux2f}. Note that for $K=K_{11}$ and $\nu=1$ we obtain
$\wtilde B\mu=0$ and so no $\mu$ term arises -- cf. \cref{susyr}.
This result though is not generic.

\subparagraph{\sl (b)} For $K=K_{22}$ and $K_{23}$, $\km$ in
\Eref{km} is both $Z$ and $\bz$ dependent and so the results in
\Eref{Bm} can be reproduced making use of the following identities
\bel
\vev{\km}&=-\frac{\sqrt{3}\no}{2\nu\sqrt{\vev{\rm g}}}\vev{\partial_Z\km}=\lm\lf\frac{4\nu^2}{3}\rg^\nu\vev{\rm g}^{1/2};\label{aux4} \\
\vev{\partial_\bz\km}&=\vev{\km}\lf\sqrt{3}/\vev{\rm g}^{1/2}-2\nu/\sqrt{3}\no\rg;\label{aux5}\\
\vev{\partial_\bz\partial_Z\km}&=\frac1\no\vev{\km}\vev{\rm
g}^{1/2}\lf\frac{8\nu^2}{3\no}\vev{\rm
g}^{1/2}-(2\nu+1)\rg.\label{aux6}\end{align}\eeqs
\\ [-0.4cm]
The derivatives of $\wkhi$ w.r.t $Z$ do not vanish only for
$K=K_{22}$ and are computed by employing \Eref{aux2f}, as
previously.

%%%%%%%%%%%%%%%%%%%%%%%%%%%%%%%%%%%%%%%%%%%%%%%%%%%%%%%%%%%%%%%%%%%%
\begin{figure}[!t]%\vspace*{-.25in}
\includegraphics[width=60mm,angle=-90]{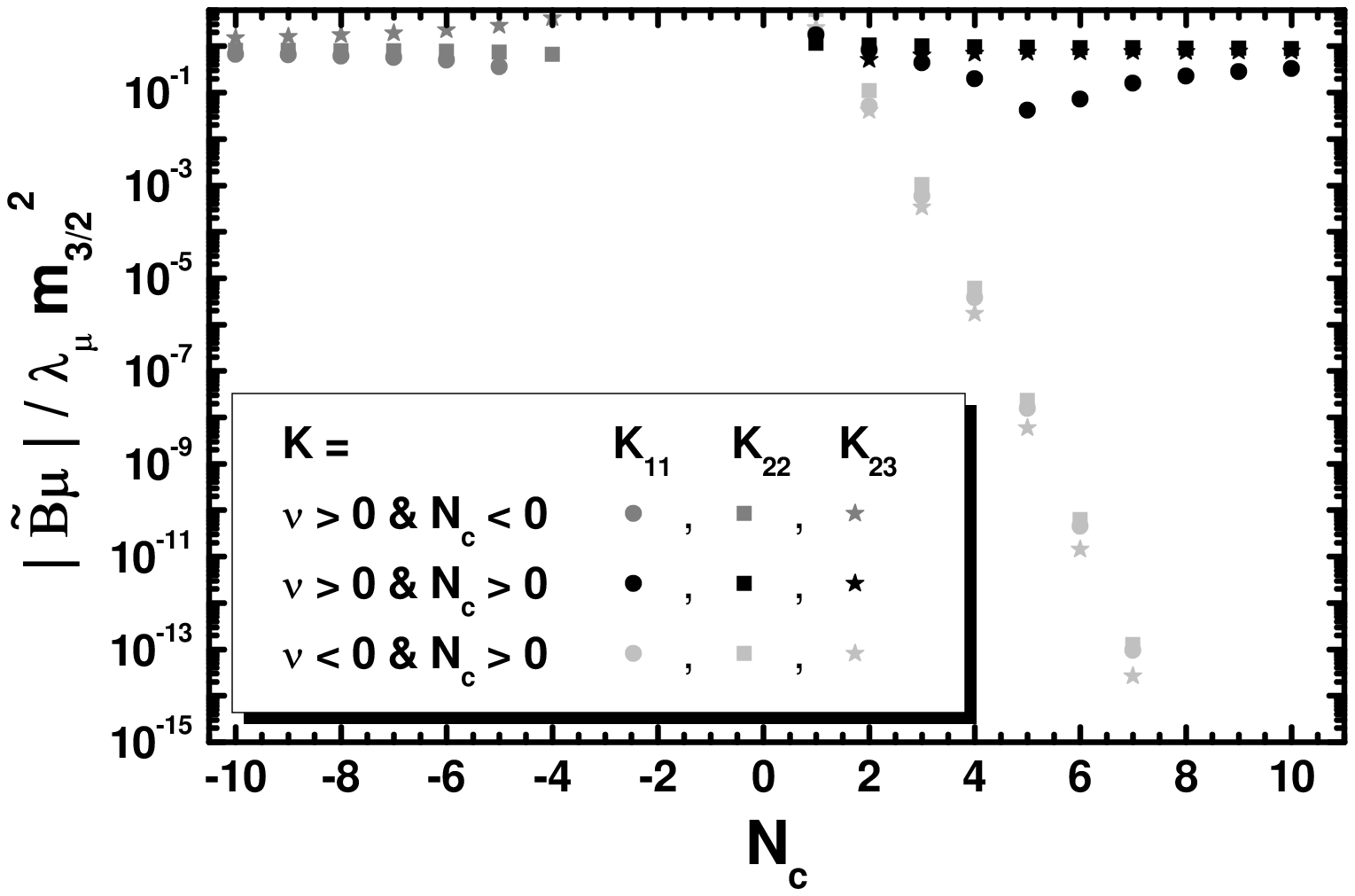}
\caption{\sl \small The ratios $\widetilde{B}\mu/\lm\mgr^2$ for
$K=K_{11}, K_{22}$ and $K_{23}$ (circles, squares and stars
respectively) versus $\no$. We take $\no<0$ (gray symbols) or
$\no>0$ and $\nu>0$ (black symbols) or $\no>0$ and $\nu<0$ (light
gray symbols).}\label{fig7}
\end{figure}
%%%%%%%%%%%%%%%%%%%%%%%%%%%%%%%%%%%%%%%%

\subparagraph{} The magnitude of the derived $\wtilde B\mu$'s
above w.r.t $\mgr^2$ is demonstrated in \Fref{fig7}, where we
present the ratios $\wtilde B\mu/\lm\mgr^2$ for $K=K_{11}$
(circles), $K_{22}$ (squares) and $K_{23}$ (stars) versus $\no$,
for integer $\no$ values and $|\no|\leq10$. We take $\no<0$ (gray
symbols) or $\no>0$ and $\nu>0$ (black symbols) or $\no>0$ and
$\nu<0$ (light gray symbols). We observe that the $\wtilde B\mu$
values, for $\lm$'s of order unity, are comparable to $\mgr^2$ for
any $\nu$ in the two first cases, whereas for $\no>0$ and $\nu<0$
these decrease as $\no$ (and $|\nu|$) increase. In his context,
therefore, besides the much heavier than $\mgr$ sfermion spectrum
-- see \Fref{fig6} -- we may obtain $\mu\ll\mgr$. The latter
result is welcome, since it enhances the electroweak naturalness
\cite{baer} of MSSM.

%We ignore here
%contributions to $M_{\rm a}$ via gauge anomalies which are treated
%as an alternative source of non-vanishing $M_{\rm a}$.

%Therefore, we can built SUSY models with
%sparticle spectrum much heavier than $\mgr$ in that case. This is
%very attractive in our days where an high-scale versions SUSY
%gains a fair amount of attention. $

\section{Conclusions} \label{con}

We completed our approach, initiated in \cref{susyr}, to the
problem SUSY breaking with a mildly violated $R$ symmetry. Our
setting is relied on the super- and \Ka s given in \eqs{whi}{khi}
respectively. We revealed that the solution $(\nu, N)=(1,-4)$
presented in \cref{susyr}  is just one from a class of solutions
obtained by considering various values for the $R$ character of
the goldstino superfield, $Z$, and letting free the geometry of
the internal space. In particular, we found ``magic" pairs $(\nu,
N)=(\nu, \no)$, with $\no$ satisfying \Eref{sol} and $\nu$ in the
domains of \Eref{all}, which allow for a naturally vanishing
cosmological constant. Specific examples are presented in
\Tref{tab1} and in Figs.~\ref{fig1} and \ref{fig2} for the cases
of $SU(1,1)/U(1)$ and $SU(2)/U(1)$ \Km s.

The presence of the cosmologically dangerous $R$ axion in the
spectrum of the model can be eluded switching on a quartic term in
the \Ka\ -- i.e., setting $k\neq0$ in \Eref{khi} -- which mildly
breaks the $R$ symmetry without modifying the SUGRA potential,
along its real direction, and the position of the Minkowski vacua
in \Eref{z0}. The mass hierarchy between the $R$ saxion and axion
and $\Gr$ is given in \eqs{mzgr}{m8gr} and deviates from the one
found in \cref{susyr}. Our scheme not only provides non-vanishing
SSB (i.e., soft SUSY-breaking) parameters, independent from $k$,
but also offers an explanation of the $\mu$ problem of MSSM
inspired by the Giudice-Masiero mechanism. Broadening the outputs
of \cref{susyr} -- according to which the SSB and $\mu$ parameters
turn out to be of the order of $\mgr$ -- here we specify patterns
of SUSY breaking which lead to values for the aforementioned
parameters  much larger or lower than $\mgr$.

The crucial difference of our proposal compared to the no-scale
models \cite{noscale18} is that here the SUSY breaking vacuum is
obtained for a specific value of $Z$, whereas in those models the
minimization of the SUGRA potential occurs along a $Z$-flat
direction, i.e., for any values of $Z$. As a consequence, $\Gr$
mass varies with $Z$ remaining thereby undetermined, whereas it
depends explicitly on $\vev{z}$ and the fundamental parameters
$\nu$ and $m$ within our models. In the majority of the no-scale
models the SSB masses turn out to be zero contrary to what happens
in our setting.

%Although we did not localize other solution $(\nu, N)$ with both
%integers besides this found in \cref{susyr}, we believe that our
%analysis is useful enough since it helps to deeply understand the
%structure of the Minkowski solutions in our models and it allows
%to generate new types of SSB parameters. E.g., zeros...

Unfortunately, our scheme does not support viable inflationary
solutions driven by $Z$ for any of the $(\nu,\no)$ values
examined. Namely, the distinct values of $\nu$ and $\no$, entailed
by Eqs. (\ref{sol}) or (\ref{nupm}), can not be reconciled with
the current inflationary data. However, the hidden sector analyzed
in our work may coexist with an inflationary one provided that
both sectors respect the same $R$ symmetry -- for
$R$-symmetry-compatible inflationary models see, e.g.,
\cref{ant,univ}. The variety of the $R$ charges of $Z$ allowed
within our framework is expected to facilitate the ``marriage'' of
both sectors. For some recent attempts towards this direction see
\cref{noscaleinfl,ant,buch,riotto,kai,linde1,ketov,king}.

Finally, it would be interesting to investigate the low-energy
consequences of the relations between the SSB mass parameters
which are usually imposed at a high scale as boundary conditions
for the evolution of the relevant renormalization group equations.
To implement this program we have to specify the SSB masses for
gauginos besides those found in \Sref{obs1} for the scalars. As
explained in \cref{susyr}, the safest option is the assumption
that the gauginos acquire SSB masses from gauge anomalies so that
the $R$-symmetry violation is conserved mild enough. This
conclusion remains intact from the modifications made in the
present work. The derived SUSY spectrum and the relevant
cosmo-phenomenological constraints could assist us to constrain
further the parameters of our models and help us to distinguish
which of them is the most compelling.

%\newpage

\paragraph*{\small\bfseries\scshape Acknowledgments} {\small I would like to thank G. Lazarides
for useful discussions. This research work was supported by the
Hellenic Foundation for Research and Innovation (H.F.R.I.) under
the ``First Call for H.F.R.I. Research Projects to support Faculty
members and Researchers and the procurement of high-cost research
equipment grant'' (Project Number: 2251).}

\appendix

\renewenvironment{subequations}{%
\refstepcounter{equation}%
% \theparentequation{\theequation}%
\setcounter{parentequation}{\value{equation}}%
  \setcounter{equation}{0}
  \def\theequation{A\theparentequation{\small\sffamily\alph{equation}}}%
  \ignorespaces
}{%
  \setcounter{equation}{\value{parentequation}}%
  \ignorespacesafterend
}
\renewcommand{\thesubsection}{{\small\sf\arabic{subsection}}}

\section{Beyond the Effective Superpotential}

The consideration of $\whi$ in \Eref{whi}, with $\nu\neq0$ a real
number, as an effective superpotential offers an elegant and
economical way for the presentation of our proposal. However, if
we confine ourselves to positive fractional $\nu$ values, we can
obtain a full-range, i.e. valid in (almost) the whole complex
plane, superpotential.

Namely, performing a field redefinition, inspired by
\cref{fractional}, we can show that our models can be formulated
equivalently without any restriction (originated from $\whi$) on
the chiral superfield. Indeed, if we define
\beq \nu=p/q~~\mbox{and}~~Z=X^q, \label{rdef} \eeq
where $p$ and $q$ are positive integers, then  $\whi$ and $\khi$
\eqs{whi}{khi} read
\beq W_{\rm H} = m X^p, \label{whi1} \eeq
which is obviously analytic in the whole complex plane, and
\beq K_{\rm H}=N\ln\big(1+\lf{|X|^{2q}-k(X^{q}-\bar
X^q)^4}\rg/{N}\big),\label{khi1} \eeq
where $|X|^{2q}\lesssim-N$ for $N<0$. The $R$ charge of $X$ is
$2/p$ and $\khi$ violates the $R$ symmetry via the $k$-dependent
term. We can easily verify that the line element $ds_{\rm H}^2$ in
the moduli space remains unaltered in this formulation of our
setting, i.e.,
\beq \label{ds1} ds_{\rm H}^2= \partial_X\partial_{\bar X} K_{\rm
H}dX d\bar X=\partial_Z\partial_\bz K_{\rm H}dZ d\bar Z,\eeq
where $\partial_Z\partial_\bz K_{\rm H}={\rm g}(Z,\bar Z)$ is
given by \Eref{ds} whereas
\beq \label{gX} \partial_X\partial_{\bar X} K_{\rm H}=q^2
|X|^{2(q-1)}{\rm g}\lf Z=X^q, \bar Z=\bar X^q\rg.\eeq
From the last expression we deduce that the $X-\bar X$ metric --
or, equivalently, the kinetic term in the $X-\bar X$ Lagrangian --
is well defined for $q\geq1$.

Repeating the analysis of \Sref{hd21} we can easily assure that
the condition in \Eref{sol}, which now takes the form
\beq \no={4p^2}/{q(3q-4p)}, \label{sol1}\eeq
allows for a SUSY-breaking Minkowski vacuum
\beq \vev{|X|}=\lf \frac{2p}{\sqrt{3}q}\rg^{1/q}~~\mbox{with}~~\im
X=0~~\mbox{and}~~\frac{p}{q}\neq\frac32\label{vevX}\eeq
-- cf. \eqs{neq32}{z0}. Some solutions to \Eref{sol1} are arranged
in the two upper lines of \Tref{tab1}. However, the present form
of the condition in \Eref{sol1} includes redundant solutions when
$p$ and $q$ are not relatively prime numbers. If we derive the
particle spectrum at the vacuum of \Eref{vevX}, we can verify that
this is identical with that exposed in Secs.~\ref{hd23} and
\ref{hd3} replacing $\nu=p/q$.  As a consequence, the supertrace
relation in \Eref{trkl} remains also intact. Finally,  the
redefinitions in \Eref{rdef} let unaffected the derivation of the
SSB parameters in \Sref{obs}.

Therefore, our construction is valid not only for the effective
$\whi$ in \Eref{whi} but also for the exact $\whi$ in \Eref{whi1}
in conjunction with $\khi$ in \Eref{khi} or  \Eref{khi1}
correspondingly.

%\newpage

\def\ijmp#1#2#3{{\sl Int. Jour. Mod. Phys.}
{\bf #1},~#3~(#2)}
\def\plb#1#2#3{{\sl Phys. Lett. B }{\bf #1}, #3 (#2)}
\def\prl#1#2#3{{\sl Phys. Rev. Lett.}
{\bf #1},~#3~(#2)}
\def\rmpp#1#2#3{{Rev. Mod. Phys.}
{\bf #1},~#3~(#2)}
\def\prep#1#2#3{{\sl Phys. Rep. }{\bf #1}, #3 (#2)}
\def\prd#1#2#3{{\sl Phys. Rev. D }{\bf #1}, #3 (#2)}
\def\npb#1#2#3{{\sl Nucl. Phys. }{\bf B#1}, #3 (#2)}
\def\npps#1#2#3{{Nucl. Phys. B (Proc. Sup.)}
{\bf #1},~#3~(#2)}
\def\mpl#1#2#3{{Mod. Phys. Lett.}
{\bf #1},~#3~(#2)}
\def\jetp#1#2#3{{JETP Lett. }{\bf #1}, #3 (#2)}
\def\app#1#2#3{{Acta Phys. Polon.}
{\bf #1},~#3~(#2)}
\def\ptp#1#2#3{{Prog. Theor. Phys.}
{\bf #1},~#3~(#2)}
\def\n#1#2#3{{Nature }{\bf #1},~#3~(#2)}
\def\apj#1#2#3{{Astrophys. J.}
{\bf #1},~#3~(#2)}
\def\mnras#1#2#3{{MNRAS }{\bf #1},~#3~(#2)}
\def\grg#1#2#3{{Gen. Rel. Grav.}
{\bf #1},~#3~(#2)}
\def\s#1#2#3{{Science }{\bf #1},~#3~(#2)}
\def\ibid#1#2#3{{\it ibid. }{\bf #1},~#3~(#2)}
\def\cpc#1#2#3{{Comput. Phys. Commun.}
{\bf #1},~#3~(#2)}
\def\astp#1#2#3{{Astropart. Phys.}
{\bf #1},~#3~(#2)}
\def\epjc#1#2#3{{Eur. Phys. J. C}
{\bf #1},~#3~(#2)}
\def\jhep#1#2#3{{\sl J. High Energy Phys.}
{\bf #1}, #3 (#2)}
\newcommand\jcap[3]{{\sl J.\ Cosmol.\ Astropart.\ Phys.\ }{\bf #1}, #3 (#2)}
\newcommand\njp[3]{{\sl New.\ J.\ Phys.\ }{\bf #1}, #3 (#2)}
\def\prdn#1#2#3#4{{\sl Phys. Rev. D }{\bf #1}, no. #4, #3 (#2)}
\def\jcapn#1#2#3#4{{\sl J. Cosmol. Astropart.
Phys. }{\bf #1}, no. #4, #3 (#2)}
\def\epjcn#1#2#3#4{{\sl Eur. Phys. J. C }{\bf #1}, no. #4, #3 (#2)}

\end{document}